\documentclass[aps,prb,reprint,superscriptaddress]{revtex4-2}
\usepackage{amsmath,slashed,amsfonts,graphicx,verbatim,siunitx}
\usepackage{xcolor}
\usepackage{graphicx}

\begin{document}
\title{Evolution from an acoustic-plasmon-mediated superconductivity to an acoustic-phonon-mediated superconductivity in bilayers}
\author{Shuyang Wang}
\author{S. Das Sarma}
\author{Jay D. Sau}
\affiliation{Condensed Matter Theory Center and Joint Quantum Institute, Department of Physics, University of Maryland, College Park, Maryland 20742, USA}
\begin{abstract}
Motivated by recent developments in van der Waals heterostructures, we revisit the acoustic plasmon mechanism of superconductivity in bilayer systems composed of a light layer (LL) and heavy layer (HL) by employing Eliashberg theory. The exchange of virtual plasmons in the HL can lead to a retarded in time attractive interaction between electrons of LL that we model through the screened interaction in the bilayer system within the random phase approximation.  We explore the evolution from acoustic plasmon mediated superconductivity to phonon mediated superconductivity by studying the evolution of $T_c$ as the HL mass is increased by a few orders of magnitude compared with the electronic mass in LL. The lower HL mass corresponds to the bilayer acoustic plasmon, while the latter regime is closer to the Born-Oppenheimer regime of acoustic phonon mediated strongly retarded pairing. The heavy HL mass limit is known to obey Migdal's theorem by virtue of the small ratio of the two individual layer masses.
We study the nonadiabatic effects for the arbitrary mass ratio with no small parameter systematically by using a frequency cut off in the Eliashberg theory, providing $T_c$ as a function of this cut off.
\end{abstract}
\maketitle

\section{Introduction}
Superconductivity, over a century after its discovery, despite having a theory to explain a majority of superconductors remains limited in temperatures to substantially below room temperature although there is no reason, in principle, for $T_c$ to be low since it is basically a coupling constant. This is quite different from other phases of matter such as metals, insulators and magnets. The BCS theory of conventional superconductivity predicts that the average phonon frequency i.e. the so-called Debye frequency controls the scale of the transition temperature of the superconductor~\cite{schrieffer2018theory}. This has motivated the question of whether  
purely electronic mechanisms based on excitions and/or plasmons can lead to superconductivity. 
The hope of achieving a higher $T_c$ through a strictly electronic mechanism arises from the fact that the electronic energy scales are typically much higher than the typical lattice energy scales of phonons, e.g., the Fermi and plasmon energies in metals are $\sim 10\,\textrm{eV}\sim 10^5 K$.
These mechanisms involve replacing the phonons in conventional superconductivity by excitons (electron-hole pairs)~\cite{little1964possibility,Allender1973} or plasmons~\cite{ihm1981demons}, both of which are very well-understood excitations. 
In fact, Little as well as Ginzburg emphasized the importance of the exciton mechanism due to its potential for achieving higher transition temperatures compared to the phonon mechanism \cite{little1964possibility,Ginzburg1970}. 
However, at present, there is no experimental evidence for the exciton mechanism of superconductivity. This is primarily because the energy gap between electrons and holes is still significant, and manipulating the density of virtual excitons remains challenging.

A simple analysis of the screened electron-electron interaction within the random-phase-approximation (RPA) suggests a strong frequency-dependent attractive interaction in a range of frequency and wave-vector space, which one could expect to lead to superconductivity. However, a recent study~\cite{sarma2025conventional} shows that an analysis of this problem within the Eliashberg framework~\cite{marsiglio2008electron} leads to a result that depends strongly on the frequency cut-off used in solving the Eliashberg integral equation. This is 
consistent with attempts to study the problem beyond the Eliashberg framework by including vertex correction~\cite{takada1993s} where the regime where the uniform electron gas is superconducting appears to depend on the precise treatment of vertex corrections.  These results can ultimately understood as a consequence of the fact that the plasmon energy for most systems is rather large leading to a break-down of Migdal's theorem~\cite{Migdal1958} as well as the retardation effect of the potential that is crucial to avoid the repulsive part of the Coulomb interaction~\cite{morel1962calculation}. The problem of high frequency plasmons can be circumvented by considering acoustic plasmons in materials with decoupled populations of electrons as has been discussed in the context of d-orbital materials~\cite{ihm1981demons}.
In fact, superconducting pairing through acoustic plasmons appears surprisingly similar to conventional superconductivity mediated by phonons when studied though the total dielectric function~\cite{Maksimovdielectric}. One caveat that will be discussed later is that the purely plasmon mechanism  ignores transverse mode (e.g. transverse acoustic phonons), which can important to superconducting pairing~\cite{allen1988total}.

Materials with decoupled populations of electrons that otherwise interact strongly via the Coulomb interaction are challenging to engineer. Recent advances in van der Waals bonded two dimensional materials such as multilayer graphene (and  2D transition metal dichalcogenides) provide a way to generate strong interlayer pairing effects with almost vanishing tunneling that can be engineered by stacking the materials at an angle~\cite{Laussy2010,Sun2021,Chou2022,sreejith2024eliashberg}. 
This arises simply from the physical fact that the electrons (and/or holes) are confined in different spatially separated 2D layers with a large potential barrier between them. In fact, multilayer graphene systems have been proposed for superconductivity related to excitons and plasmons such as those mediated by exciton-polaritons~\cite{Sun2021}, acoustic plasmons~\cite{FatemiRuhman,millisplasmon} and exciton condensates~\cite{Laussy2010}. 
On the other hand, as is apparent from studies of plasmon mediated superconductivity~\cite{sarma2025conventional}, despite the apparent applicability of RPA, a careful analysis of cut-off dependence is necessary to assess the likelihood of acoustic plasmon mediated superconductivity. 

\begin{figure}
    \centering
    \includegraphics[width=0.95\linewidth]{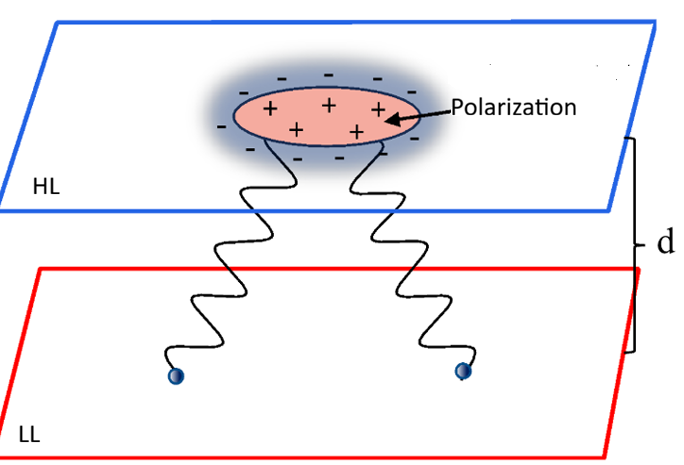}
    \caption{System schematic: a bilayer system consisting of a  lighter 2DEG (LL) with mass $m_1$ and a heavier 2DEG (HL) with mass $m_2$. The mass $m_1=0.023 m_e$ (where $m_e$ is the electron mass) is chosen to match that of bilayer graphene. The mass $m_2$ will be chosen to vary from $m_1<m_2<100 m_1$ to interpolate from the symmetric case to phonon mediated pairing. The distance between the two 2DEGs is $d=10.5\AA$. The inter-layer tunneling is assumed to vanish e.g. because of momentum mismatch. Polarization  provided by the HL and induce the attractive pairing interaction of electrons in the LL. This is essentially acoustic plasmon mediated pairing in the LL. }
    \label{fig:1}
\end{figure}

In this work, we leverage the total dielectric function framework~\cite{Maksimovdielectric,allen1988total}, which essentially uses the Eliashberg framework with the RPA screened Coulomb interaction, to provide an approach to circumvent the challenges of calculating superconductivity mediated by acoustic plasmons. The central idea is to vary the effective mass of the carriers that provide the plasmons. Tuning to the regime where this mass is large simulates pairing of electrons by ions, which are much heavier than electrons, as in the well-understood regime of conventional superconductivity. 

 When the interlayer mass ratio is very large, then the heavy mass layer can be thought of as made of ions  with heavy mass, and the resulting acoustic plasmon then is nothing other than the acoustic phonon of the heavy mass layer, with the system obeying Migdal's theorem by virtue of the large mass ratio.
One can then study how the superconductivity is modified by going to smaller "ion" masses as well as different frequency cut-offs. More specifically,  we will study superconductivity in a bilayer heterostructure comprised of a light layer (LL) with a smaller effective mass $m_1$ and a heavier layer (HL) with a larger effective mass $m_2$ ( Fig.~\ref{fig:1}) . The idea here is that our results should be exact in the very karge heavy mass layer limit since the superconductivity is now mediated by acoustic phonons whereas the results will depend systematically on the frequency cut off, due to the importance of vertex corrections, as the two masses approach each other with the resulting superconductivity being mediated by the bilayer acoustic plasmons~\cite{sarmamadhukar,hwangsarma}.  We will be interested in the superconducting properties of the LL electrons, while the HL electrons will play the role analogous to ions in a conventional superconductor. The van der Waals bonding structure allows us to place the electrons in the two layer at a relatively small distance $d\sim 3.4\AA$ from each other, which is much smaller than the typical separation of electrons $\sim 10\,$nm in these materials. Consequently, the LL and HL electrons can effectively be considered to be in the same material. Furthermore, varying the electric field in a dual gated device and/or twist angle in a heterostructure allows one to vary both the mass $m_2$ as well as the density $n_2$ in the layer HL providing control of the acoustic plasmon frequency. The Fermi energy of the superconducting layer LL can similarly be varied by a gate voltage. Therefore, the bilayer system shown in Fig.~\ref{fig:1} is an ideal system to explore the cross-over from conventional superconductivity in the large $m_2$ region to the smaller $m_2$ region that is natural for acoustic plasmons. This latter region is where one might hope to see higher $T_c$ superconductors, in analogy with superconducting hydrides. It should be noted that we do not expect the entire parameter regime of mass and densities to be accessible in a single dual gated device. Rather the device in Fig.~\ref{fig:1} is a conceptual device that allows one to identify parameter regimes with promising superconducting behavior. The central concern in this approach would be the choice of the frequency cut-off $\omega_c$~\cite{sarma2025conventional}, which is analogous to the Coulomb-pseudopotential in Eliashberg theory~\cite{marsiglio2008electron}. But as we will discuss, one can obtain reasonable estimates of this cut-off based on the properties of the isolated LL layer as well as superconductivity in the large $m_2$ limit.

This paper is organized as follows. In Section \ref{sec:2}, we provide a detailed description of how the RPA screened Coulomb interaction in the LL-HL heterostructure shown in Fig.~\ref{fig:1} leads to acoustic plasmon mediated pairing. 
In Section \ref{sec:3}, we describe the Eliashberg gap equation, as well approximation and cutoff schemes~\cite{sarma2025conventional} that we use in our numerical results. In Section~\ref{sec:4} we present numerical results and discuss the implications for the critical temperature $T_c$ for acoustic plasmon based superconductivity. Finally, in Section~\ref{sec:5}, we summarize our findings and discuss various implications and caveats.

\section{Pairing interaction for the bilayer system}\label{sec:2}
We examine a multilayer configuration in which a 2DEG  (HL, $l=2$) with a heavy mass $m_{l=2}$ is placed on top of a 2DEG  (LL, $l=1$)  with a lighter mass $m_{l=1}$ (Fig. \ref{fig:1}). The ratio $m_2/m_1$ is the main tuning parameter in the theory of superconductivity described in the current work. The layers are assumed to have densities $n_{l=1,2}$ and are separated by a distance $d$, so that the only  interaction between the electrons is the intra-layer and inter-layer Coulomb interaction. We assume zero interlayer tunneling of electrons. Such a pair of metallic layers supports an acoustic plasmon mode in addition to the conventional plasmon. Each layer is assumed to be a single band 2D electron system with isotropic effective mass.

 Here, we focus on the pairing of LL electrons generated by acoustic plasmon mechanism (i.e. neglecting lattice phonons).
The effective electron-electron pairing interaction in LL resulting from acoustic plasmons in the system can be written in terms of the RPA dielectric function $\varepsilon(q,\omega)$ as: 
\begin{equation}\label{eqn:Veff_ex}
    V_{eff}(q,\omega)=\left [\varepsilon^{-1}(q,\omega) V(q)\right]_{11},
\end{equation}
 where 
\begin{equation}
    \varepsilon(q,\omega)=1-\Pi(q,\omega)V(q),
\end{equation}
is the bilayer dielectric function, $q$ is the interaction wave-vector and $\omega$ is the imaginary frequency. In the above, $V(q)$ represents the Coulomb potential:
\begin{equation}
    V(q)=\left(
     \begin{array}{cc} V_1(q) & V_2(q)\\V_2(q)&V_1(q)\\ 
     \end{array}
     \right),
\end{equation}
with intra-layer ($V_1$) and inter-layer ($V_2$) Coulomb potentials given by:
\begin{equation}
    V_1(q)=\frac{2\pi e^2}{\kappa q},\quad V_2(q)=\frac{2\pi e^2}{\kappa q}e^{-d q}
\end{equation}
, where $\kappa$ is the background dielectric constant.
The polarization $\Pi$ of the bilayer is a diagonal matrix:
\begin{equation}
    \Pi(q,\omega)=\left(
     \begin{array}{cc} \Pi_1(q,\omega) & 0\\0&\Pi_2(q,\omega)\\ 
     \end{array}
     \right),
\end{equation}
where $\Pi_{1,2}(q,\omega)$ are the polarization functions of LL and HL, respectively. Combining the matrix forms of the polarizability $\Pi$ and the Coulomb interaction $V$ with Eq.~\ref{eqn:Veff_ex}, we can write the spectral function of the pairing interaction in LL, $U(q,\omega)=N_1(0)V_{eff}(q,\omega)$, as 
\begin{multline}
    U(q,\omega)
 =N(0)
 \frac{V_1-\Pi_2(V_1^2-V_2^2)}{(1-V_1\Pi_1)-\Pi_2\{V_1-(V_1^2-V_2^2 )\Pi_1\}}\label{eq:Ufinal}
\end{multline}
where $N_1(0)$ is the  density of states of the LL (i.e. $l=1$). Note that the $(q,\omega)$ dependence of the RHS has been suppressed for notational convenience.
This effective screened interaction can be written in terms of the intra-layer and inter-layer screened Coulomb interactions $W_{1,2}=V_{1,2}/(1-\Pi_1 V_1)$ as: 
\begin{align}
    &U(q,\omega)=N_1(0)W_1+\frac{\Pi_2 N_1(0) W_2^2}{1-W_3\Pi_2},
\end{align}
where $W_3=W_1-\frac{(V_1^2-V_2^2)\Pi_1}{1-V_1\Pi_1}$. The first contribution to $U(q,\omega)$ in the above form is the intra-layer screened interaction, which is repulsive for much of the range of frequency. The latter term proportional to $\Pi_2(1-W_3\Pi_2)^{-1}=(\Pi_2^{-1}-W_3)^{-1}$ is the acoustic plasmon-induced attractive interaction with the renormalized acoustic plasmon propagator $(\Pi_2^{-1}-W_3)$. This expression formally matches the conventional form of a Boson-induced pairing interaction~\cite{marsiglio2008electron} once we identify $W_2$ as the electron-Boson interaction. 

The scaled interaction $U(q,\omega)$ can be written in a dimensionless form in terms of dimensionless momenta  $z=q/2 k_{F,l}$ and frequency  $u=\omega/q v_{F,l}$. 
 For example $U(q,\infty)=N_1(0)V_1(q)=r_s/z\sqrt{2}$, where $r_s$ is the dimensionless 2D electron gas parameter of the layer LL, $k_{F,l}$ is the Fermi wave-vector and $v_{F,l}$ the Fermi velocity of the layer $l$. 
 The 2D electron gas parameter $r_s$ is defined for 
$r_s =(\pi n_1)^{-1/2}/a_B$, where $a_B=\hbar/e^2m_1$ is the Bohr radius. 
Similar to the interaction, the scaled polarization functions for the LL can be written as \cite{Stern1967,sreejith2024eliashberg}: 
\begin{multline}
\varpi_j(q,\omega)\equiv N_j(0)^{-1}\Pi_j(q,\omega)\\
=-(2 z)^{-1}[2z-f(z-iu)-f(z+iu)],    
\end{multline}
where $f(w)=w\sqrt{1-w^{-2}}$ and the square root is defined as the branch with the positive real part.
Since the pairing interaction in Eq.~\ref{eq:Ufinal} acts on electrons in the layer LL, the polarizability $\varpi_2$ needs to be rescaled using the identities $N_2/N_1=m_2/m_1$, $k_{F,2}/k_{F,1}=(n_2/n_1)^{1/2}$ and $v_{F,2}/v_{F,1}=(m_1/m_2)(n_2/n_1)^{1/2}$.

Next, we consider superconductivity arising only from the dynamically screened Coulomb interaction considered here, and work in the jellium approximation with no lattice phonons. But, for $m_2/m_1\gg 1$, the heavy layer acts as an effective ionic layer for the light layer with the acoustic plasmon becoming the effective acoustic phonon for the bilayer as in the original Bardeen-Pines model of electron-phonon interaction in metals~\cite{bardeenpines}.

\section{Superconducting gap equation}\label{sec:3}
The momentum and frequency dependent pairing interaction $U(q,\omega)$ can lead to a superconducting pairing gap that can be calculated using the Eliashberg equation~\cite{marsiglio2008electron}. For $s-$wave superconductivity, one can replace the pairing interaction by its Fermi-surface average~\cite{marsiglio2008electron} over the LL Fermi surface:
\begin{align}
&U(\omega) =N_1(0)(-\pi)^{-1}\int_{FS}dk dk'U(k-k';\omega).
\end{align}
The formalism we use here closely follows the framework 
for studying plasmon mediated superconductivity discussed in the recent work Ref.~\onlinecite{sarma2025conventional}. Specifically, it was noted that the long-range Coulomb interaction $V(q)$ can lead to infrared divergences in $U(\omega)$. To avoid this, we introduce a cut-off $k_c$ so that the Fermi-surface average is taken slightly away from the Fermi surface so that $|k-k'|>k_c$. As discussed in recent work~\cite{sarma2025conventional}, the most reasonable choice of $k_c$ corresponds to $k_c\sim T_c/v_{F,1}$, where $T_c$ is the superconducting transition temperature~\cite{sarma2025conventional}.

\begin{figure*}
    \centering
    \includegraphics[width=0.95\linewidth]{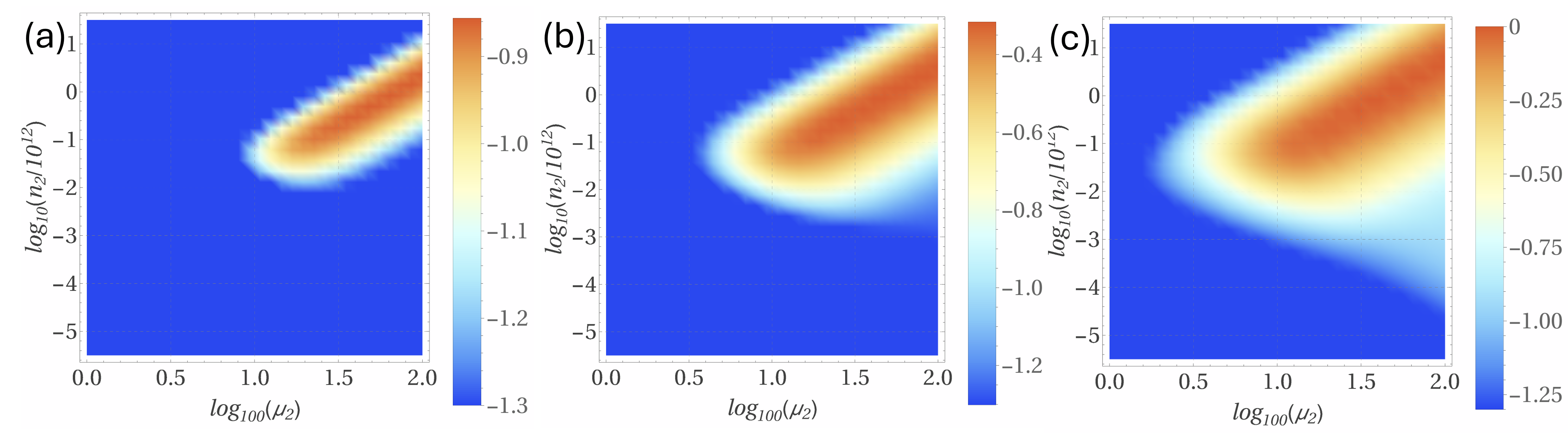}
    \caption{(Color online) $Log_{10}(T_c (\textrm{in K}))$ for superconductivity in the bottom layer of in Fig.~\ref{fig:1} with density $n_1=10^{12}cm^{-2}$ as a function of the density $n_2$ and mass $m_2=\mu_2 m_1$ of the top layer.  The cut-off $\omega_c$ is chosen to be  
     $\omega_c=0.15 E_{F1}, 0.18 E_{F,1},0.19 E_{F,1}$ respectively from left to right. The momentum cut-off is chosen to be $k_c=0.25\times 10^{-2} k_{F,1}$. With these parameters the LL layer by itself has a transition temperature $T_{c1L}<50\,$mK. As explained in the text the cut-off $\omega_c$ is determined by vertex corrections, which then determines the strength of superconductivity, similar to the Coulomb pseudopotential~\cite{marsiglio2008electron}. The SC $T_c$ increases as mass increases towards the limit consistent with phonon-mediated SC and also with decreasing density as the acoustic plasma frequency decreases. The range of significant $T_c$ shifts to higher $m_2$ and lower $n_2$ as $\omega_c$ decreases. }
    \label{fig:2}
\end{figure*}

Since we focus on weak coupling SC, we ignore quasiparticle renormalization in the Eliashberg equation, so that the linearized gap equation~\cite{marsiglio2008electron} 
has the form of a frequency dependent BCS equation:
\begin{align}
\phi_m=-\pi  T \sum_{m'}\frac{U(\omega_m-\omega_{m'})}{|\omega_{m'}|}\phi_{m'},\label{eq:Eliash}
\end{align}
where $\omega_m=(2m+1)\pi T$ are the discrete imaginary Matsubara frequencies at temperature $T$. 
 Following recent work on plasmon mediated SC~\cite{sarma2025conventional}, in order to avoid vertex-correction effects, following Migdal's theorem~\cite{Migdal1958}, we restrict the sum to $|\omega_m|<\omega_c$. As shown in prior work~\cite{sarma2025conventional}, the choice of the frequency cut-off $\omega_c$ is formally similar to estimating the Coulomb pseudo-potential in Eliashberg theory  for the regular phonon-mediated metallic superconductivity~\cite{marsiglio2008electron}.   The above equation can be converted to a symmetric eigenvalue problem by rescaling the gap function to $\psi_m=\phi_m/\sqrt{\omega_m}$. The gap equation for $\psi_m$ is a symmetric eigenvalue problem
\begin{align}
\psi_m=-\pi  T \sum_{m'}\frac{U(\omega_m-\omega_{m'})}{\sqrt{|\omega_{m'}||\omega_{m}|}}\psi_{m'},\label{eq:Eliashsymm}
\end{align}
where the transition temperature $T=T_c$ if the lowest eigenvalue matches $-1$.

We now discuss in more detail the choice of the cut-off frequency $\omega_c$. As is clear from previous work~\cite{sarma2025conventional}, this issue is more subtle in the case of plasmon-mediated pairing, which does not have a clearly defined Debye frequency. Because of this, we solve the above Eliashberg equation, which includes both attractive and repulsive interactions, numerically with the introduction of a cut-off $\omega_c$ that restricts the frequencies  $|\omega_m|<\omega_c$ in Eq.~\ref{eq:Eliashsymm}. Here, we argue how this can lead to a bound on $T_c$ if $\omega_c$ is chosen to be small, i.e., $\omega_c\ll E_{F,1}$ to avoid vertex corrections~\cite{Migdal1958}. For $\omega_c$ small enough, Migdal theorem applies allowing the neglect of vertex corrections. To understand this we note that the linearized Eliashberg equation Eq.~\ref{eq:Eliash} without the frequency cut-off $\omega_c$ is identical to the instability equation for the superconducting instability $\phi(\omega_m)\equiv \phi_m$ provided the interaction kernel $U(\omega)$ is replaced by the appropriately momentum space averaged four-fermion vertex~\cite{tsai2005renormalization}. However, this interaction vertex in general is complicated to compute e.g. by a renormalization group~\cite{tsai2005renormalization} since it includes vertex corrections. On the other hand, based on Migdal's theorem~\cite{Migdal1958}, the interaction kernel $U(\omega)$ for low frequencies $|\omega|<\omega_c$ for small $\omega_c\ll E_{F,1}$ receives negligible vertex corrections. Therefore, the low-frequency part of the interaction $U(|\omega|<\omega_c\ll E_{F,1})$ is approximately correct.
This allows us to solve  Eq.~\ref{eq:Eliashsymm} to generate a lower bound on $T_c$ even without being able to calculate the high-frequency part of the kernel $U(|\omega|>\omega_c)$. This is done by solving Eq.~\ref{eq:Eliashsymm} with the constraint $|\omega_{m,m'}|<\omega_c$.
 In more mathematical detail, the symmetric form of the linearized Eliashberg equation where $T_c$ is related to the lowest eigenvalue can be viewed as a variational problem since the lowest eigenvalue of this eigenvalue equation can be obtained by minimizing the expectation value of the scaled interaction matrix in Eq.~\ref{eq:Eliashsymm}. Thus, the lowest eigenvalue and correspondingly $T_c$ can be bounded from below by solving Eq.~\ref{eq:Eliashsymm} with a cut-off, even if the high frequency part of $U$ cannot be calculated due to vertex corrections. Thus, a frequency cut-off allows a systematic investigation of $T_c$ avoiding the issue of the importance of vertex corrections and the inapplicability of Migdal's theorem using a single parameter, $\omega_c$.

 Since we do not estimate the vertex correction at a given cut-off point $\omega_c$, the results presented in this work depend on an unknown parameter that we must estimate. The situation is similar to conventional electron-phonon superconductivity where the Coulomb pseudopotential (arising from the direct electron-electron repulsion) $\mu^*$~\cite{marsiglio2008electron} is unknown. It should be noted that although this is not a significant impediment in many conventional superconductivity calculations, where $\mu^*$ is believed to be between $0.1-0.2$, van der Waals systems are very different from three dimensional metals where this number has not been estimated and the value of $\mu^*$ in these systems is unknown. One can bound $\omega_c$ using constraints on superconductivity in related systems. In this work, we choose $\omega_c$ to be near, but smaller than the value where the LL system is superconducting in isolation. This is motivated by the fact that there is no evidence of superconductivity in  two dimensional electron gases  with single layer (e.g. monolayer graphene) pure Coulomb interactions, though it has been suggested in some theoretical calculations~\cite{takada1993s,cai2022superconductivity}. In our numerical results, we will find that reducing $\omega_c$ below this value will reduce $T_c$ as expected from the arguments made above, but does not affect $T_c$ qualitatively in the regime of lower-acoustic plasmon as expected from comparison with the case of electron-phonon coupling. Since the choice of the cut-off point $\omega_c$ is substantially below $E_{F,1}$, which is where vertex corrections are significant, we expect the resulting $T_c$ presented in this work to be a reasonable estimate.   In the next section, we describe the results of the numerical solution of the above equation.

\section{Numerical Results for $T_c$}\label{sec:4}
We now discuss the dependence of the critical temperature resulting from solution of the superconducting gap equations for acoustic plasmons in the configuration shown in Fig.~\ref{fig:1}. Motivated by van der Waals systems the interlayer distance $d$ between the layers LL and HL is chosen to be  $d=10.5\textup{\AA}$ together with a background dielectric constant $\kappa=2.5$ (i.e. corresponding to graphene on $SiO_2$ as the substrate material). The effective mass of the LL layer is chosen to match bilayer graphene $m_1=0.04$ (relative to the electron mass). Motivated by the availability of different effective masses in van der Waals heterostructures, we will study the dependence of $T_c$ with the heavier effective mass $m_2$ of the layer HL. Furthermore, we will also study the dependence on the densities of both layers $n_{1,2}$ in the vicinity of the density $n_{1,2}\sim 10^{12}/cm^2$, which is typical in these devices.  Changing $m_2$ with fixed $m_1$ is equivalent to changing the dimensionless mass ratio $m_2/m_1$.

As can be seen in Fig.~\ref{fig:2}, the system shown in Fig.~\ref{fig:1} for appropriate ranges of density and effective mass manifests acoustic plasmon mediated SC with $T_c$ comparable to $\sim 1\,K$. (Of course, explicit values of $T_c$ depends on the choice of the parameters, but the trends as functions of mass, density, and frequency cut off are independent of parameter choice.)  The dominant superconductivity occurs in the range of large $m_2/m_1>100$, where one can view the superconductivity as related to electron phonon superconductivity with the electrons in HL serving the role of ions in a conventional superconductor~\cite{allen1988total} and the acoustic plasmons essentially becoming effective emergent acoustic phonons for the system. As expected, the superconducting $T_c$ is suppressed at small density $n_2$, where the $T_c$ should match the $T_c$ of the isolated LL layer. Consistent with prior single-layer work~\cite{sarma2025conventional}, $T_c$  approaches a finite small value $T_{c1L}\lesssim 50\,$ mK depending on the choice of the frequency cutoff $\omega_c$. In the case of panel (a), the $50\, mK$ is likely simply an upper bound on $T_c$. 
The two panels Fig.~\ref{fig:2} show superconducting transition temperatures $T_c$ for the frequency cutoff values $\omega_c=0.15 E_{F1}$ and $\omega_c=0.2 E_{F,1}$ respectively.
The cutoffs are chosen with the constraint that the corresponding $T_c$s for the isolated LL  are below  $T_{c1L}\lesssim 50\,$mK, which is consistent with  bilayer graphene by itself not being  known to exhibit superconductivity. The representative value of $k_c$ used for the plot is adjusted to accommodate the higher $T_c$ for panel (c). Consistent with previous work~\cite{sarma2025conventional} $T_c$ is found to increase with decreasing $k_c$, so we are using the $k_c$ corresponding to the largest value of $T_c$ in a plot so as to underestimate $T_c$ away from the maximum. Thus, while Fig.~\ref{fig:2} provides an estimate of the maximum $T_c$ that can be expected from the acoustic plasmon mechanism, it does not provide a minimum since $\omega_c$ (which is effectively determined by vertex corrections) could be much smaller than $\omega_c\sim 0.15 E_{F,1}$ leading to an insignificant $T_c$. Unfortunately, as discussed in the previous section, the cut off dependence is unavoidable since there is no effective way for including vertex corrections to all order as necessary in the absence of the Migdal theorem for an arbitrary mass ratio between the layers. This is formally similar to the Coulomb pseudo-potential dependence of $T_c$ in conventional electron-phonon superconductivity~\cite{marsiglio2008electron}.

\begin{figure}
    \centering
    \includegraphics[width=1.05\linewidth]{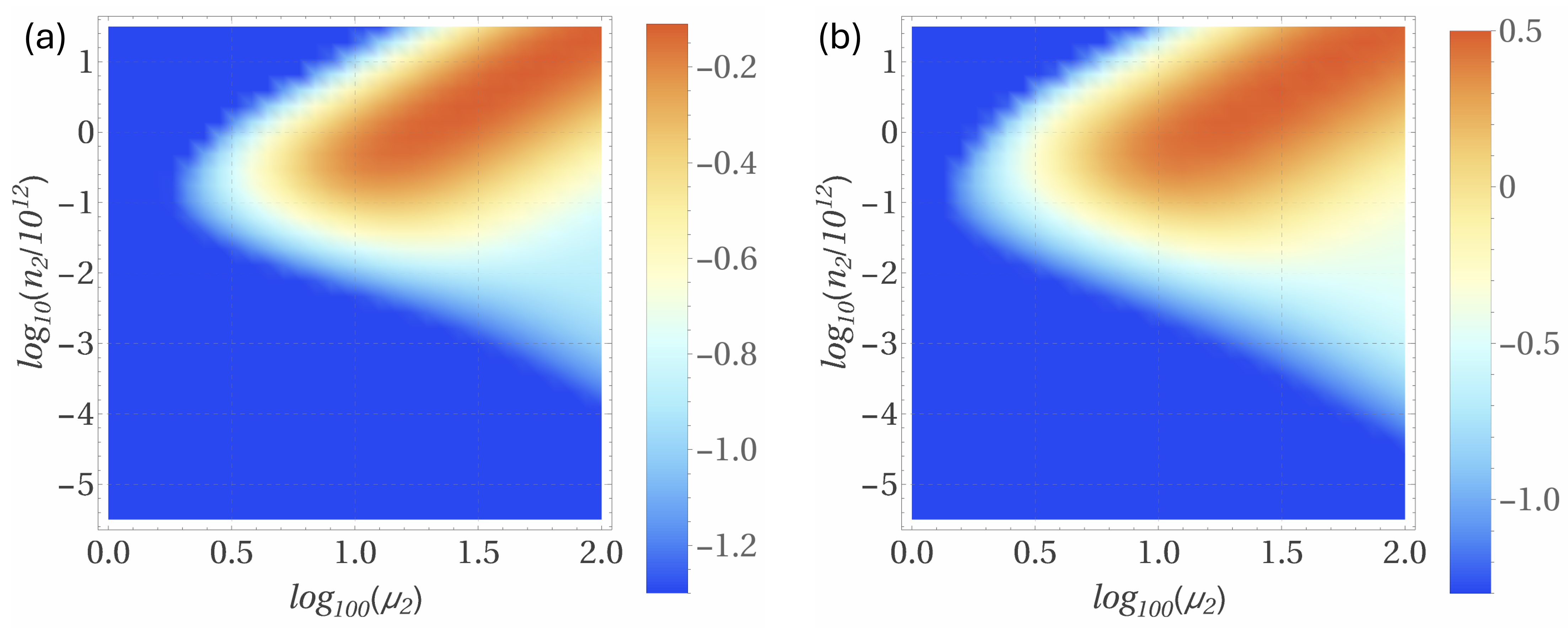}
    \caption{(Color online) $Log_{10}(T_c (\textrm{in K}))$ for superconductivity in LL with density $n_1=4\times 10^{12}cm^{-2}$ as a function of the density $n_2$ and mass $m_2=\mu_2 m_1$ of the top layer similar to Fig.~\ref{fig:2}. The figure shows that the $T_c$ in panel (b) is enhanced by lowering the distance between the layers in panel to $d=5.25\AA$ relative to the the distance in panel (a), which is as indicated in Figs.~\ref{fig:1},~\ref{fig:2}. 
    The cut-off $\omega_c$ is chosen to be  
     $\omega_c=0.21 E_{F,1}$. The momentum cut-off is chosen to be $k_c=0.25\times 10^{-2} k_{F,1}$. With these parameters the LL layer by itself has a transition temperature $T_{c1L}<50\,$mK. As explained later in the text the cut-off $\omega_c$ is determined by vertex corrections and the dependence on $\omega_c,k_c$ is explored in a later plot. The SC $T_c$ increases as mass increases towards the limit consistent with phonon-mediated SC and also with decreasing density as the acoustic plasma frequency decreases. The range of significant $T_c$ shifts to higher $m_2$ and lower $n_2$ as $\omega_c$ decreases. }
    \label{fig:3}
\end{figure}

One of the central questions that this work intended to address is whether the effect of lowering the mass $m_2$, which should effectively raise the Debye frequency, can lead to a high temperature superconductor. The results in Fig.~\ref{fig:2} show that while this works for a range of mass as well as density, the $T_c$ in this case is suppressed when the effective acoustic plasmon frequency approaches the cut-off. Thus, the extent to which lowering $m_2$ can increase $T_c$ is limited in this case by non-adiabatic vertex correction effects, where the Migdal theorem may not apply by virtue of the effective phonon frequency being large or the effective Fermi energy being low. The acoustic plasmon frequency is suppressed at smaller $n_2$ or larger $m_2$ and is seen to suppress the corresponding $T_c$. Smaller density $n_2$ also reduces the strength of the pairing interaction, which then can completely eliminate $T_c$ (within the Eliashberg framework) below some critical value. Therefore, we find that the acoustic-plasmon mediated superconductivity does not lead to high $T_c$ despite the additional control of the Debye frequency provided by the set-up in Fig.~\ref{fig:1}.  
This leads to the question of how to possibly compare these results to the superconductivity in the hydrides~\cite{hydride}, where an enhanced Debye frequency is believed to be the driver for high $T_c$. An answer to this is provided by the results in Fig.~\ref{fig:3}, which are similar to Fig.~\ref{fig:2}, except that the density $n_1$ of the LL is increased and the distance $d$ is lowered in panel Fig.~\ref{fig:3}(b). We find that increasing the density (relative to Fig.~\ref{fig:2}) while reducing the inter-layer distance $d$ can significantly increase the $T_c\sim 4\,$K. This suggests that if one could reach densities that are a few orders of magnitude higher and separations $d$ that are an order of magnitude smaller, with a smaller dielectric constant, one could increase $T_c$ to a range comparable with the hydride materials~\cite{hydride}. However, this is clearly not a possibility in a van der Waals material where the distances are limited by van der Waals bonding and the densities are limited by the gate capacitance. We emphasize that we are not suggesting any connection between our theory and hydride superconductivity, where a superconducting mechanism is not yet established in the literature. Instead we are only using the  hydride superconductivity as a motivation to ask why increasing the effective Debye frequency does not lead to increased $T_c$.

\section{Discussion and conclusion}\label{sec:5}
In this paper, motivated by recent developments in van der Waals materials, 
we have revisited the acoustic plasmon mechanism for superconductivity in a LL-HL bilayer platform by varying the effective mass and effective carrier density in the layers. Layered 2D materials provide a possibility of placing a gas of electrons (HL in Fig.~\ref{fig:1}) that provide the pairing glue in close proximity to the potential superconducting carrier layer (LL in Fig.~\ref{fig:2}). In addition, such van der Waals heterostructures allow, in principle, to tune the system over a large range of densities and effective masses. We have theoretically studied the superconducting $T_c$ over a large range of densities and effective masses, without specifying the precise device to generate this.  The calculation of superconducting $T_c$ is done using a combination of Eliashberg theory and dynamically RPA screened Coulomb interaction. Based on earlier work~\cite{sarma2025conventional} using the same framework on plasmon mediated superconductivity, we have carefully accounted for the dependence of the calculated $T_c$ on the frequency cut-off $\omega_c$, which plays the role of the unknown Coulomb pseudopotential. The frequency cut off also enables a careful incorporation of the possible importance of unknown vertex correction since the frequency range above the cut off is, by definition, the high frequency range which Migdal's theorem allows us to ignore. The central result of this work is that, while acoustic plasmon mediated superconductivity is possible in these systems, the transition temperature is likely smaller that $T_c\lesssim 1\,$K for any values of the layer masses and layer densities, thus invalidating the hope of a high$-T_c$ superconductor induced by a plasmon mechanism. This is somewhat surprising, since one might have expected the ability to increase the effective Debye frequency in the large mass ratio limit might have enhanced the transition temperature as in superconducting hydrides~\cite{hydride}.  We find that the inter-layer separation $d$ as well as the electron-density become limiting factors in the bilayer geometry in Fig.~\ref{fig:1} even though the effective "ionic" mass $m_2$ is no longer a limit. Thus, the physics is complicated, and depends crucially on all the relevant physical parameters sensitively: layer effective masses, layer densities, and layer separation.
 
There are several aspects of conventional electron-phonon superconductivity that were ignored by the theory in this work. LL-HL system only allows longitudinal excitations i.e. acoustic plasmons to participate in pairing. This is different from phonon mechanisms where both longitudinal and transverse phonons can participate. This distinction has been shown to be relevant even within the dielectric formalism for phonon mediated superconductivity where the transverse coupling is mediated by  Umklapp scattering~\cite{allen1988total}. It is possible that the coupling to such transverse modes can appear in purely electronic systems when the HL system enters a Wigner crystal regime. In addition, the RPA interaction considered here can also account for interactions mediated by intra-layer excitons leading to an excitonic pairing mechanism~\cite{little1964possibility,Ginzburg1970}. Both of these interactions are screened differently by the LL electrons and might increase $T_c$ significantly from the estimates purely from acoustic plasmons. On the experimental front, it is possible that these discussions are relevant to superconducting variants of more recently developed bulk layered Moire structures~\cite{checkelskyMoire} but much more work using more realistic band structure would be necessary to figure out the role of plasmons in 2D moire superconductivity.

The use of RPA, valid only in the high-density limit $(r_s<<1)$ is also questionable at arbitrary densities, but there are no controlled systematic many-body approximation beyond RPA which is useful in the current work.  In spite of all the simplifying approximations, which are all generically used in theories of electronic materials, the key qualitative finding of our work should survive:  The system should manifest superconductivity as $m_2/m_1$ is tuned continuously from very large values to unity since the basic Eliashberg theory should be valid for $m_2/m_1\gg 1$ where the superconductivity arises from an acoustic phonon type pairing to $m_2/m_1 \sim 1$, where the superconductivity arises from a acoustic plasmon pairing (and $T_c$ could be in mK range) since the same mode makes a continuous crossover from a bosonic acoustic plasmon mode for $m_2/m_1\sim 1$ to an acoustic phonon type boson for $m_2/m_1\gg 1$ with no obvious phase transition between the two regimes.  We conclude that superconductivity may be generic in bilayers with large mass difference whereas it may be drastically suppressed when the layer mass difference is small.

\section*{Acknowledgement}
This work is supported by the Laboratory for Physical Sciences through its support of the Condensed Matter Theory Center. J.S. and S. W. acknowledges support from the Joint Quantum Institute.

\bibliographystyle{apsrev4-2}
\bibliography{bibl} 
\end{document}